\documentclass[12pt]{article}
\usepackage{amsmath,amssymb}
\usepackage[dvips]{graphicx}
\usepackage{cite}

\def\Label#1{\label{#1}%
  \smash{\hbox to0pt{\raise1ex\hbox{\tiny[#1]}\hss}}}

\newcommand{\be} {\begin{equation}}
\newcommand{\ee} {\end{equation}}
\newcommand{\bea} {\begin{eqnarray}}
\newcommand{\eea} {\end{eqnarray}}

\newcommand{\eq}[1]{(\ref{#1})}

\newcommand{\pat}{\partial}

\def\half{\frac{1}{2}}

\newcommand\apple{{$\overline {\textrm{D}}(p+1)\ $}}

\begin{document}

\rightline{HIP-2008-39/TH}

\vskip 1cm \centerline{\large {\bf Winding effects on brane/anti-brane pairs}}

\vskip 1cm

\renewcommand{\thefootnote}{\fnsymbol{footnote}}
\centerline{{\bf Niko Jokela,$^{1,2}$\footnote{najokela@physics.technion.ac.il} Matti J\"arvinen,$^{3}$\footnote{mjarvine@ifk.sdu.dk}
and Sean Nowling$^{4,5}$\footnote{sean.nowling@helsinki.fi} }}
\vskip .5cm
\centerline{\it ${}^{1}$Department of Physics } \centerline{\it
Technion, Haifa 32000, Israel}
\centerline{\it ${}^{2}$Department of Mathematics and Physics } \centerline{\it
University of Haifa at Oranim, Tivon 36006, Israel}
\centerline{\it ${}^{3}$ Center for High Energy Physics }
 \centerline{\it University of Southern Denmark,
Campusvej 55, DK-5230 Odense M, Denmark}
\centerline{\it ${}^{4}$Helsinki Institute of Physics } \centerline{\it
P.O.Box 64, FIN-00014 University of Helsinki, Finland}
\centerline{\it ${}^{5}$Department of Mathematics and Statistics } \centerline{\it
P.O.Box 68, FIN-00014 University of Helsinki, Finland}

\setcounter{footnote}{0}

\renewcommand{\thefootnote}{\arabic{footnote}}

\begin{abstract}

We study a brane/anti-brane configuration which is separated along a compact direction by
constructing a tachyon effective action which takes into account transverse scalars.
Such an action is relevant in the study of HQCD model of Sakai and Sugimoto of chiral
symmetry breaking, where the size of the compact circle sets the confinement scale.
Our approach is motivated by string theory orbifold constructions and gives a route to model
inhomogeneous tachyon decay.
We illustrate the techniques involved with a relatively simple
example of a harmonic oscillator on a circle.
We will then repeat the analysis for the Sakai-Sugimoto model
and show that by integrating out the winding modes will provide us with a renormalized action
with a lower energy than that of truncating to zero winding sector.

\end{abstract}

\newpage




\section{Introduction}


Brane/anti-brane configurations play an important role in many applications of string theory.
A particular example is the Sakai-Sugimoto model for chiral symmetry breaking in holographic QCD (HQCD) \cite{Sakai:2004cn,Sakai:2005yt}. 
In this setting one has pointlike D8-branes and anti-D8-branes on a Neveu-Scherk circle.
The brane/anti-brane pairs are separated asymptotically in the radial direction on the circle. For some minimal value of the radial coordinate they are connected, thus providing a geometrical description of chiral symmetry breaking.

In the original proposal the model only describes massless chiral fermions.
It has been a challenging task to study the quark mass deformation of the model.
To be able to add the mass to the fermions, it is crucial to take into account the bi-fundamental field
of the open string, the tachyon, which comes from strings stretching between D8-$\overline{\mathrm{D}}8$-branes.
Since, according to the usual holographic dictionary, by turning on the expectation value of the tachyon one finds, that the normalizable mode of this field corresponds to the quark condensate while the non-normalizable
mode gives rise to the mass of the quarks \cite{Bergman:2007pm,Dhar:2007bz,Dhar:2008um}.

In the case of Sakai-Sugimoto model one has pointlike D8-$\overline{\mathrm{D}}8$-branes on a circle and is forced
to deal with infinitely many
winding strings which wrap the compact space. As discussed in \cite{Taylor:1996ik}, after
T-duality, exciting winding modes correspond to inhomogeneities in field configurations
wrapping the compact space.\footnote{See also \cite{Sugimoto:2004mh}.}
It is therefore interesting to study what kind of finite size effects would we see by condensing winding
tachyon fields in the Sakai-Sugimoto model.
In particular, as we shall discuss below, one should notice that the size of the circle enters nontrivially into the D8-$\overline{\mathrm{D}}8$-brane effective
action. In the Sakai-Sugimoto model, the radius of the circle determines the confinement scale of the dual gauge theory. This relationship thus constitutes one of our main motivations to study physics associated with this parameter.

In non-compact spacetimes there are known constructions for the effective actions of brane/anti-brane pairs both in flat and curved spacetimes \cite{Kraus:2000nj,Takayanagi:2000rz,Jones:2002sia,Sen:2003tm,Sen:2004nf,Garousi:2004rd,Garousi:2007fn}. Generically, these take the form of a Dirac-Born-Infeld (DBI) action which is modified to include tachyon modes.  When discussing small fluctuations, it is a simple exercise to apply the methods discussed in \cite{Taylor:1996ik} to build actions describing brane/anti-brane pairs which are either pointlike or wrap a compact space.  In the case of branes wrapping the compact space, inhomogeneities in the tachyon field lead to many winding tachyons after a Kaluza-Klein reduction.  In the T-dual picture of pointlike branes, this structure is realized through an infinite order shift orbifold.\footnote{In \cite{Taylor:1996ik} there is a problematic issue of the action's normalization. We shall comment on this issue for the case of many winding tachyons in Subsection \ref{sec:TDBIcirc}.} Due to the broken supersymmetry one expects a complicated potential coupling the tachyons to the open string moduli. For pointlike branes/anti-branes on a circle with radius $R$, the effective potential, at the quadratic level, from strings stretching between the branes takes the form
\be\label{eq:schematic}
 V_{S^1}\sim \sum_w -\frac{|T_w|^2}{2\alpha'}+|T_w|^2\frac{\left(\ell+2\pi R w\right)^2}{(2\pi\alpha')^2} \ .
\ee
The sum in (\ref{eq:schematic}) is over the infinite number of winding strings on the circle. If the value of the moduli field $\ell$ is large with respect to the string length, the tachyons are all extremely massive. As the length field is decreased, the tachyons become massless and then tachyonic. For string scale brane separations, one expects to lose the open string description as the tachyons condense \cite{Sen:2004nf}.

Typically, one simplifies the analysis by considering turning on the tachyons one at a time. Then, if the circle's radius is much larger than the string length, it is anticipated that one could simply truncate to the zero winding tachyon. When thinking about examples with a finite number of tachyons, one would expect the zero winding tachyon to condense long before the higher winding tachyons become massless. The corresponding brane/anti-brane system would be effectively non-compact.  For other approaches to tachyon condensation on compact spaces, see \cite{Sen:2003zf,Terashima:2008jz}.

However, in \cite{Sen:2002vv,Larsen:2002wc,Lambert:2003zr,Balasubramanian:2004fz,Jokela:2005ha} it is suggested that inhomogeneous decay is important in flat space and one would like to investigate its possibility in compact spaces.  Unfortunately, the truncation to zero winding simply does not allow for tachyon inhomogeneities.  Logically, there is another possibility that might invalidate a simple truncation of the full action.  In the presence of infinitely many degrees of freedom, nontrivial scaling limits might be found, which need not satisfy the usual intuitions.  Also, one still must deal with an infinite number of degrees of freedom.
We begin by considering a tachyon profile with large inhomogeneities.
We will employ a non-trivial scaling limit to develop an effective field theory description of the brane/anti-brane configuration in compact space. Being an effective field theory, this new Lagrangian will involve only a finite number of fields, yet retain knowledge about the compact nature of the problem.
Although counter intuitive, we shall argue that this scaling solution should be used whenever it has a lower energy than that of the single tachyon background. We show that this is the case in the Sakai-Sugimoto model, see \cite{Bergman:2006xn,Bergman:2007wp,Bergman:2007pm,Bergman:2008sg,Bergman:2008qv,Edalati:2008xr,Johnson:2008vna,Aharony:2006da,Parnachev:2006dn,Antonyan:2006vw, Aharony:2008an, Hashimoto:2008sr,Dhar:2007bz,Dhar:2008um,Casero:2007ae,Hashimoto:2008zw,Hata:2007mb,McNees:2008km,Davis:2007ka,Erdmenger:2007cm,Thompson:2008qw,Seki:2008mu,Argyres:2008sw,Nawa:2006gv}
for recent studies in this direction.

The outline of this paper is as follows. In Section \ref{sec:ppbar}, we shall review tachyon effective actions in compact spaces when the branes are both extended and pointlike along a circle.  In Section \ref{sec:oscil} we use a shift orbifold to describe a harmonic oscillator on a circle. This simple classical mechanical problem has similar winding dependence as the brane examples. We will be forced to introduce a regulator to define the theory even at the classical level due to infinitely many degrees of freedom. After identifying a scaling solution we shall be led to a $V_{eff}\sim\cos(\ell/R)$ renormalized potential, as is natural for particles coupled by an oscillator on a circle. After this simple warm up we shall pursue a field theory realization of these ideas in Section \ref{sec:Gar} which is relevant for an effective low-energy description to string theory. Starting from an effective action for separated pointlike branes and anti-branes on a circle, we will use an orbifold to study the winding effects. In Section \ref{sec:SS} we use this new effective
Lagrangian in the Sakai-Sugimoto HQCD model for chiral symmetry breaking, allowing for a non-vanishing open string tachyon field. Such models were considered in \cite{Casero:2007ae,Bergman:2007pm,Dhar:2007bz,Dhar:2008um}. Studying the system numerically, we find that the scaling solution is slightly preferred by the energetic considerations. In the final Section \ref{sec:conc} we summarize and discuss possible outgrowths of this work.

\section{ D$(p+1)$-\apple Wrapping a Circle \label{sec:ppbar}}

To motivate to study winding tachyons associated with separated
D$p$-$\overline{\mathrm{D}}p$-brane pair it is instructive to
switch to the
equivalent T-dual picture, where the geometrical separation is
transformed to an open string
gauge field strength background of the parallel
D$(p+1)$-$\overline{\mathrm{D}}(p+1)$-brane pair \cite{Taylor:1996ik}.
Although, the D$(p+1)$-brane configuration would be more complicated to study we will find out
that considering winding tachyons for D$p$ system is quite natural.

We are interested in studying only one pair of D$(p+1)$-branes in which case
the worldvolume action is Abelian.
For a small (complex) tachyon
expectation value, $|T|$, the Lagrangian for coincident D$(p+1)$-$\overline{\mathrm{D}}(p+1)$-brane pair reads \cite{Sen:2003tm}
\be
\label{eq:firstL}
 \mathcal{L} \sim
-\frac{1}{4}\left(F^{(1)}_{\mu\nu}\right)^2-\frac{1}{4}\left(F^{(2)}_{\mu\nu}\right)^2-
|\left(\partial_\mu - i(A_\mu^{(1)}-A_\mu^{(2)})\right)T|^2 +
\frac{1}{2\alpha'}|T|^2\ +\ldots\ .
\ee
We have also ignored higher derivative corrections and transverse
scalars. The last term in (\ref{eq:firstL}) comes from the potential $V(|T|)$ and its coefficient is fixed
such that it matches with the worldsheet calculations. Different authors use different potentials \cite{Sen:2004nf,Garousi:2004rd,Garousi:2007fn} but all agree at small $|T|$, which is the regime of interest to us in this paper. 
It is more natural to write (\ref{eq:firstL})
in terms of the relative gauge field, $A^- = A^{(1)}-A^{(2)}$.

If the worldvolume coordinate, $x^1$, is compactified on a circle of radius $\tilde{R}$, it is natural to introduce mode expansions. Following \cite{Taylor:1996ik}, we have
\bea
 A^{(-) 1}(x,x^1)  & = & \sum_{n=-\infty}^\infty X^{(-)1}_n(x) e^{ i n x^1/\tilde{R}} \\
 T(x,x^1)          & = & \sum_{n=-\infty}^\infty \tau_n(x) e^{i nx^1/\tilde{R}}\ .
\eea

Allowing for a nontrivial gauge field only in the $x^1$ direction, the part of the Lagrangian (after integrating over $x^1$) which involves tachyons is rewritten with these new variables as
\bea\label{eq:tach}
 \mathcal{L} & \sim & -\sum_{n, m=-\infty}^\infty  \tau_n\bar{\tau}_m \left(   \left(\left(\frac{n}{\tilde{R}}\right)^2-\frac{1}{2\alpha '}\right)\delta_{m,n} - \frac{n+m}{\tilde{R}}X^{(-)1}_{m-n}\right)\nonumber \\
 & & \ \ \ \ \ \ \ \ \ - \sum_{n,m,\ell\ =-\infty}^\infty \tau_n\bar{\tau}_m X^{(-)1}_\ell X^{(-)1}_{m-n-\ell} \ .
\eea

It is readily apparent that it is difficult to consider tachyon inhomogeneities. If one considers a profile for $T(x,x^1)$ which is sharply peaked in the $x^1$ direction, many $\tau_n(x)$ are excited. That is, for large inhomogeneities, one must confront infinitely many winding tachyons, $\tau_n(x)$.  For such configurations it would be useful to have an effective description which involves fewer degrees of freedom, but still captures the winding physics necessary to describe inhomogeneous backgrounds.  To this end we shall find it conceptually simpler to describe the T-dual theory of pointlike D$p$-$\overline{\mathrm{D}}p$-branes on a circle of radius $R=\frac{\alpha '}{\tilde{R}}$.  The Wilson line, $\int dx^1 A^{(-)}_1$, T-dualizes into a relative separation between the D$p$-branes along the new compact direction.

The terms in (\ref{eq:tach}) T-dualize into
\bea
 \mathcal{L} &\sim&  -\sum_{n, m=-\infty}^\infty  \tau_n\bar{\tau}_m   \left( \left(\left(\frac{R n}{\alpha'}\right)^2-\frac{1}{2\alpha '}\right)\delta_{m,n} - \frac{R (n+m)}{\alpha'}X^{(-)1}_{m-n}\right)\nonumber \\ && \ \ \ \ \ \ \ \ \ - \sum_{n,m,\ell\ =-\infty}^\infty \tau_n\bar{\tau}_m X^{(-)1}_\ell X^{(-)1}_{m-n-\ell} \ .
\eea
These describe open strings connecting the D$p$-$\overline{\mathrm{D}}p$-branes.  As in \cite{Taylor:1996ik}, the above action also has an interpretation in terms of a shift orbifold acting on a single brane/anti-brane pair in flat space.  The fields $\tau_n(x)$ are interpreted as winding tachyons. In the language of \cite{Taylor:1996ik}, $X^{(-)1}_n$ for $n\neq 0$ are related to off-diagonal entries in the embedding matrix for D$p$-brane positions.  As such, they correspond to a noncommutative embedding. Because we are interested in description with pointlike branes, we shall assume
\be
 X_n^{(-)1} = 0\quad \forall n\neq 0 \ .
\ee
In the original D$(p+1)$ system this corresponds to considering only constant gauge fields along the circle, $A^{(-)1}(x,x^1) = A^{(-)1}(x ) $.  This assumption may break down when the D$p$-branes separation distance becomes string scale.

Even assuming that the D$p$ position matrix is diagonal, one is left with the daunting task of understanding the behavior of infinitely many winding tachyon modes.  At this point, the most common approach is to truncate the tachyon Lagrangian to only include zero winding modes.  This is usually justified by imagining that the tachyons condense one at a time.  In such cases, it is true that as the branes approach one another, the zero winding tachyon's mass becomes negative before any of the other tachyons.  There is a caveat to this argument, it is based upon experience with finite numbers of fields.  Another issue is that such an analysis would not even allow for the possibility of inhomogeneous decay of the D$(p+1)$ system.

Rather than simply truncating the Lagrangian by hand, one could instead try to systematically integrate out the winding tachyons. If such effective Lagrangian can be shown to be energetically favorable, one should take it seriously, in spite of its counterintuitive character.  While this might be a nice story, for it to be useful one must still confront the infinite number of fields in the Lagrangian. One natural approach would be to look for effective field theories written in terms of a few degrees of freedom which capture the correct qualitative features of the many winding modes.  In what follows, this is the strategy that we shall follow.  We will introduce a highly localized tachyon profile $T$, which will naively lead to divergences in the classical action.  As in other cases, dealing with the naive divergences will lead us to a renormalization of the naive action.  Although the renormalization of a classical expression is unusual, it has a familiar origin in the presence of infinitely many degrees of freedom.

To consider large inhomogeneities we shall take all winding tachyons to be equal, $\tau_n(x) \equiv \tau(x)$. The Lagrangian in (\ref{eq:tach}) becomes
\be
 \mathcal{L} \sim -\sum_{n=-\infty}^\infty |\tau(x)|^2\left(\left(\frac{R n}{\alpha'}+ X^{(-)1}_0(x)\right)^2-\frac{1}{2\alpha'}\right)\ .
\ee
To compare with \cite{Garousi:2004rd} we shall define a field $X^{(-)1}_0(x) = \frac{\ell(x)}{2\pi \alpha'}$.  Assuming vanishing gauge fields in the D$p$-brane worldvolume, we find
\be\label{eq:tach2}
S =-\sum_{n=-\infty}^\infty T_{p}\int d^{p+1}x\ \left(  2\pi\alpha'|\partial_\mu \tau|^2+ \frac{1}{2}\left(\partial_\mu \ell \right)^2+ |\tau(x)|^2 \left(\frac{ \left(2\pi R n+ \ell(x)\right)^2}{2\pi\alpha'}-\pi\right)\right)\ .
\ee

As mentioned above, the theory in (\ref{eq:tach2}) is manifestly ill-defined,
and requires some form of regularization.  In the next Section, we shall develop
the necessary formalism in a simple harmonic oscillator example before
returning to the D$p$ and D$(p+1)$ theories.


\section{Oscillators on a circle}\label{sec:oscil}

Consider the classical theory of a one-dimensional harmonic oscillator connecting two particles, at positions $X^1$ and $X^2$.  As in Section \ref{sec:ppbar}, let us attempt to study this system, constrained to live on a circle of radius $R$, through a shift orbifold. Although this example may seem too trivial, we shall see that it is sufficient to display the typical difficulties encountered when trying to implement the proposal we discussed in the Section \ref{sec:ppbar}.

We separate the center-of-mass motion from the relative coordinate
\be
 \mathrm{X} = \frac{X^1+X^2}{\sqrt{2}}\quad ; \quad \ell = \frac{X^1-X^2}{\sqrt{2}} \ .
\ee
In what follows, the center-of-mass coordinate $\mathrm{X}$ always decouples, allowing us to write a theory for $\ell$ only.  For an oscillator on the real line we have the potential
\be
 V_{single}(\ell) = \frac{k}{2}\ell^2 \ .
\ee

We would like to build an effective theory for oscillators on a circle. When working on the circle we have new possibilities: the oscillator may wrap the circle multiple times before attaching to the particles. Such winding modes are incorporated by working on the covering space and implementing the method of images,
\be
 V_{S^1}\sim\sum_{n=-\infty}^\infty V_{single}(\ell+2\pi R\ n) \ .
\ee
In effect, this implements an orbifold where one would like to identify the fields $\ell\sim\ell+2\pi R n$.

We need to give a meaning to the following (Euclidean) Lagrangian, $\theta \equiv \frac{\ell}{R}$,
\be\label{sumosc}
 \mathcal{L} = \frac{mR^2}{2}\sum_{n=-\infty}^\infty\left(\dot\theta^2+\frac{k}{m}\left(\theta+2\pi n\right)^2 \right) \ .
\ee
Although classical, the infinite collection of modes require a regularization to sensibly
define the theory. As in more familiar field theory examples, we shall introduce a cut-off parameter, $\epsilon$, to render the effective action finite. Next we will introduce renormalized quantities through wavefunction and coupling constant counterterms. As we remove the cut-off, the divergences must be absorbed into the definition of the counterterms as functions of $\epsilon$. As in other cases, it is not sufficient to simply absorb the divergences. We must also satisfy renormalization conditions. For us it is most natural to impose conditions as $R\rightarrow \infty$ to ensure that the correct non-compact theory is obtained. As we shall see, the main subtlety lies in the form of the finite terms necessary to implement the renormalization conditions.

\subsection{Cut-off regularization}

The above action (\ref{sumosc}) has two obvious features: it diverges and it is invariant under $\theta\rightarrow \theta+2\pi$. We want to preserve this invariance in the renormalized action. When considering the type of effective action we could define, we note the following facts. In any renormalization scheme, the leading divergence will be independent of fields. We also expect divergent subleading terms which may involve $\dot{\theta}$, but there can be no finite order polynomials of $\theta$ due to the symmetry.

To extract physical information, we will look for scaling solutions. These solutions are required to respect the shift symmetry and satisfy certain renormalization conditions. In this case, the renormalization conditions amount to specifying the renormalized action's form at small renormalized field values
\be\label{eq:rencond}
 \mathcal{L} \sim \frac{m R^2}{2}\left(\dot{\theta}^2+\frac{k}{m}\theta^2\right) \ .
\ee
Condition (\ref{eq:rencond}) is tantamount to demanding the correct $R\rightarrow\infty$ behavior.

\paragraph{Regulator:}

To isolate the divergences, let us introduce a cut-off regulator\footnote{In Appendix \ref{app:anal} we use an alternate regulator based upon an analytic continuation to obtain the same regulated effective action. Even using a cut-off scheme, there is some ambiguity in how the theory is regulated. For instance, because the bare coefficients are $\epsilon$-dependent, there are ordering ambiguities associated with the position of the derivative, $\partial_\epsilon$. One finds that each specific choice leads to different values for the counterterms, but the renormalized Lagrangian is always the same.} for Lagrangian (\ref{sumosc}),
\bea
\mathcal{L}_{\epsilon}  & = & \frac{m R^2}{2}\left(-\frac{\partial\ }{\partial \epsilon}\right)\left[\sum_{n=-\infty}^\infty e^{-\epsilon\left(\dot{\theta}^2+ \frac{k}{m}\left(\theta+2\pi n\right)^2 \right)}\right]\\
\mathcal{L}             & = & \lim_{\epsilon\rightarrow 0^+}\mathcal{L}_\epsilon\ .
\eea

This expression should be reorganized to isolate the divergences. Using a Jacobi theta function, $\left(\Theta_3 (\nu|\tau ) = \sum_{n=-\infty}^\infty e^{\pi i \tau n^2} e^{2\pi i \nu n}\right)$,
\bea
\mathcal{L}_{\epsilon} & = & \frac{m R^2}{2}\left(-\frac{\partial\ }{\partial \epsilon}\right)\left[ e^{-\epsilon\left(\dot{\theta}^2+ \frac{k}{m}\theta^2 \right)}\Theta_3\left(i2\epsilon\theta\frac{k}{m}\Big{|}i4\pi \epsilon \frac{k}{m}\right)\right]\\
                       & = &\frac{m R^2}{2}\left(-\frac{\partial\ }{\partial \epsilon}\right)\left[ e^{-\epsilon\left(\dot{\theta}^2 \right)}\sqrt{\frac{m}{4\pi k}}\frac{1}{\epsilon^{1/2}}\Theta_3\left(\frac{\theta}{2\pi}\Big{|}\frac{im}{4\pi k \epsilon}\right)\right] \ . \label{eq:cutlag}
\eea

As expected, we see that there is a field-independent divergence. In addition, there is a subleading divergence proportional to $\dot{\theta}^2$. All the $\theta$ dependence is locked in exponentially vanishing terms. Without looking for scaling solutions, all the $\theta$ dependence simply drops out of the effective action, making it impossible to satisfy the renormalization condition (\ref{eq:rencond}) (to obtain the correct decompactification limit). To proceed, we shall introduce suitable multiplicative renormalization factors which will kill the divergent terms and soften the convergence of the $\theta$-dependent terms, as is necessary to satisfy the renormalization condition (\ref{eq:rencond}).

\paragraph{Counterterms:}

After isolating the divergences, we will introduce various renormalization factors to absorb them. It is important to note that one cannot simply add polynomial counterterms due to the shift symmetry:
\bea
 Z_\ell    & : & \ell_0 = Z_\ell^{1/2}\ell \\
 Z_R       & : &  R_0 = Z_R R \\
 Z_k       & : &  k_0= Z_k k \\
 Z_\Lambda & : &  \mathrm{vacuum\ energy\ shift} \ .
\eea

Writing the Lagrangian (\ref{eq:cutlag}) as a function of renormalized quantities, we find
\bea
\mathcal{L}_{\epsilon} &=& \frac{m Z_R^2 R^2}{2}\left(-\frac{\partial\ }{\partial \epsilon}\right)\left[ e^{-\epsilon\frac{Z_\ell}{Z_R^2}\left(\dot{\theta}^2\right)}\sqrt{\frac{m}{4\pi k}}\frac{1}{(Z_k\epsilon)^{1/2}}\Theta_3\left(\frac{\theta}{2\pi}\frac{Z_\ell^{1/2}}{Z_R}\Big{|}\frac{im}{4\pi k Z_k\epsilon}\right)\right]+ Z_\Lambda \ . \nonumber
\eea

It turns out that we can uniquely fix the counterterms by requiring the correct decompactification limit.\footnote{The effective action is uniquely determined up to a discrete choice determining the winding sector in which the field $\theta$ lives. This ambiguity is fixed by requiring that the smallest shift leaving the action invariant is $\theta\rightarrow\theta+2\pi$. Higher winding sectors would only need to be invariant under shifts $\theta\rightarrow\theta+\frac{2\pi}{w}$ for a winding number $w$. Fixing $w=1$ may simply be considered a fixed condition necessary to define a renormalization scheme.} The shift symmetry of the renormalized action fixes
\be\label{eq:rel1}
 Z_\ell =   Z_R^2 \  .
\ee
Demanding that the kinetic term for $\theta$ is correctly normalized fixes
\be\label{eq:rel2}
   Z_R^4 = \frac{4\pi k}{m} Z_k\epsilon \ .
\ee

Using relations (\ref{eq:rel1}) and (\ref{eq:rel2}), we can rewrite the Lagrangian as
\bea
 \mathcal{L}_\epsilon = \frac{m R^2}{2}\dot\theta^2 + \frac{m R^2}{2 } (Z_k\epsilon)^{1/2}\left(-\frac{\partial\ }{\partial\epsilon}\right)\left[\frac{1}{(Z_k\epsilon)^{1/2}} \Theta_3\left(\frac{\theta  }{2\pi} \Big{|}\frac{i m}{4\pi k} \frac{1}{Z_k\epsilon}\right)\right]+Z_\Lambda \ .
\eea

All that remains is to solve for $Z_k$ and $Z_\Lambda$. Notice, if $Z_k$ is only a polynomial in $\epsilon$, all of the $\theta$ dependence will still vanish exponentially due to the properties of $\Theta_3$. Therefore, we need to soften this exponential behavior into a simple powerlaw. Focusing on the $\theta$ term, we must obtain $\frac{k R^2}{2}\theta^2$ for small $\theta$. Equivalently, we must demand
\be \label{eq:thetacond}
 -\frac{\partial\ }{ \partial\epsilon} \Theta_3\left(\frac{\theta  }{2\pi}\Big{|}\frac{im}{4\pi k}\frac{1}{Z_k\epsilon}\right) \sim \frac{k}{m}\theta^2\ \ \ (\mathrm{for\ small\ }\theta) \ .
\ee
While satisfying (\ref{eq:thetacond}) may look complicated, we can solve it with a simple ansatz, $\frac{m}{4k Z_k\epsilon} = -\ln(c\cdot\epsilon^p)$.
Equation (\ref{eq:thetacond}) becomes
\bea
    -\frac{\partial\ }{ \partial\epsilon}\left(1+2\sum_{q=1}^\infty\cos( q\theta)\left(c \epsilon^p \right)^{q^2}  \right) \sim \frac{k}{m}\theta^2.
\eea
Choosing
\be
    Z_k = -\frac{m}{4k}\frac{1}{\epsilon\ln\left(\frac{ k}{m}\epsilon\right)} \ ,
\ee
we find a finite term in the Lagrangian, $-\left(\frac{m R^2}{2}\right)\frac{2k}{m}\cos{\theta}+ \mathcal{O}(\epsilon^2)$. When Taylor expanded, this includes the term $\frac{kR^2}{2}\theta^2$, as desired.

Finally, we should determine the vacuum renormalization constant. Using the solution for $Z_k$, the field-independent terms in the action are (in the small field limit)
\bea\label{eq:vaccond}
 \frac{mR^2}{2 } (Z_k\epsilon)^{1/2}\left(-\frac{\partial\ }{\partial\epsilon}\frac{1}{(Z_k\epsilon)^{1/2}}\right) -kR^2  + Z_\Lambda \ .
\eea
Our renormalization condition forces (\ref{eq:vaccond}) to vanish, which fixes $Z_\Lambda$ in terms of all the other parameters.

\paragraph{Summary:}

For convenience, we summarize the above counterterms:
\bea
 Z_\ell      & = & Z_R^2  \\
   Z_R^4       & = & \frac{4\pi k}{m}\epsilon Z_k \\
 Z_k         & = & -\frac{m}{4k}\frac{1}{\epsilon\ln\left(\frac{k}{m}\epsilon\right)} \\
   Z_\Lambda   & = & kR^2+\frac{m R^2}{2}(Z_k\epsilon)^{1/2}\frac{\partial\ }{\partial\epsilon}\frac{1}{(Z_k\epsilon)^{1/2}}\ .
\eea

Plugging in the wavefunction renormalization factors, we find the renormalized effective action
\bea
 \mathcal{L}_{ren} & = &\lim_{\epsilon\rightarrow 0} \frac{m Z_R^2 R^2}{2}\!\left(-\frac{\partial\ }{\partial \epsilon}\right)\!\!\left[ e^{-\epsilon\frac{Z_\ell}{Z_R^2}\left(\dot{\theta}^2\right)}\!\sqrt{\frac{m}{4\pi k}}\frac{1}{(Z_k\epsilon)^{1/2}}\Theta_3\!\left(\frac{\theta}{2\pi}\frac{Z_\ell^{1/2}}{Z_R}\Big{|}\frac{im}{4\pi k Z_k\epsilon}\right)\!\right] +  Z_\Lambda \nonumber\\
             & = & \frac{mR^2}{2}\dot{\theta}^2 +  kR^2 \left(1-\cos( \theta)\right)\\
             & = & \frac{m}{2}\dot{\ell}^2 + kR^2\left(1-\cos\left(\frac{\ell}{R}\right)\right)\ .
\eea

After introducing a scaling solution, we found the action for two point particles on a circle with a spring connecting them.\footnote{Another place where these scaling solutions might be useful is in $2+1$ dimensional lattice compact QED.  For compact QED a common lattice Lagrangian is the Wilson Lagrangian, $\mathcal{L}_W\sim 1-\cos F$.  Although it manifestly respects the compact nature of the gauge group, it can be difficult to use due to its nonlinear character.  Using our results, it might be easier to use a Gaussian Lagrangian, $\mathcal{L}\sim \sum_n (n+F)^2$, which one now knows is in the same universality class.  }
 In principle, we could carry out this analysis for theories whose non-compact descriptions contain higher order interactions. These interactions would simply induce higher harmonics in the effective potential with coefficients tuned to achieve the correct $R\rightarrow \infty$ limit.


\section{Tachyon effective action}\label{sec:Gar}

In the previous Section we developed an effective action for a classical mechanical system by integrating out winding modes.  We would now like to return to the D$p$-$\overline {\textrm{D}}p$ (and D$(p+1)$-$\overline {\textrm{D}}(p+1)$) systems.  Our starting point will be the action for a brane/anti-brane pair separated in the $x^1$ direction.  In fact, it is trivial to embed the branes into curved space
\be\label{eq:curvedaction}
 \mathcal{L} = - \mu_{p}e^{-\varphi}\sqrt{-g} \left(2+ g_{11}(\partial \ell)^2+ 2\pi \alpha' |\partial \tau|^2+g_{11}\frac{|\tau|^2\ell^2}{2\pi\alpha'}-\pi|\tau|^2\right)\ .
\ee
In (\ref{eq:curvedaction}) it is assumed that $g_{11}$ does not depend on $\ell$.
The $\sqrt{-g}$ is simply the measure from the bulk metric, restricted to the worldvolume coordinates; we have chosen to explicitly write all the $\partial\ell$ dependence. Also, the field $\ell$ is the component transverse to the branes. Finally, the first term represents the standard volume term present when expanding DBI actions.

\subsection{Tachyon DBI on a circle}\label{sec:TDBIcirc}

Using the effective action for branes separated on a non-compact space, we wish to implement a shift orbifold in the $x^1$ direction.  Assuming the same inhomogeneous tachyon profile for $T$, the Lagrangian for D$p$-$\overline{\mathrm{D}}p$ becomes
\bea
    \mathcal{L}_W \sim -\sum_{n=-\infty}^\infty \!\! \mu_{p}e^{-\varphi}\!\sqrt{-g}\!\left(2+g_{11}(\partial \ell)^2+ 2\pi \alpha' |\partial \tau|^2+g_{11}\frac{|\tau|^2(\ell+2\pi R_4 n)^2}{2\pi\alpha'}-\pi|\tau|^2\right)\ .\nonumber
\eea

As discussed in Section \ref{sec:ppbar}, experience with field theories of a finite number of fields may be misleading.  We will look for a simpler effective action which retains some of the winding physics.  We will find it natural to adjust $\mu_{p}$ when we sum over winding contributions. In the Section \ref{sec:SS} we will provide an explicit example of such a model where the scaling solution is shown to be energetically favorable.

Following the harmonic oscillator example, we introduce a regulator for the winding physics
\bea\label{eq:regtbi}
    \mathcal{L}_{W,\epsilon} &=& -\sum_{n=-\infty}^\infty e^{-\varphi}\sqrt{-g}\nonumber\\
    &&\left(2+ 2\pi \alpha' |\partial \tau|^2-\pi|\tau|^2+g_{11}(\partial \ell)^2-\frac{g_{11}  |\tau|^2 }{2\pi\alpha'}\partial_\epsilon \right)\mu_{p} e^{-\epsilon(\ell+2\pi R n)^2} \ .
\eea
As is usual, there is some ambiguity in how the theory is regulated. When using different regulators we find that the precise form of the counterterms must be adjusted. However, the physical quantity, the renormalized Lagrangian, will be invariant under all these choices.

Using properties of theta functions, (\ref{eq:regtbi}) can be rewritten as
\bea
\mathcal{L}_{W,\epsilon} & = & -e^{-\varphi}\sqrt{-g}\left(2+2\pi\alpha' |\partial\tau|^2-\pi|\tau|^2 +g_{11}(\partial \ell)^2-\frac{g_{11}|\tau|^2}{2\pi \alpha' }\partial_{\epsilon}\right)\nonumber\\
& &\ \ \ \ \ \ \ \ \ \ \ \times \mu_{p}\frac{1}{\sqrt{4\epsilon R^2 }}\Theta_3\left(\frac{\theta}{2\pi}\Big{|}\frac{i}{4\pi \epsilon R^2 }\right)
\ .\eea

The scaling solution is very similar to the harmonic oscillator example. We will let $\mu_{p}$ and $ R $ to be functions of the cut-off $\epsilon$.\footnote{Rescaling $\mu_{p}$ can be thought of as rescaling the string coupling constant $g_s$. An alternate perspective on this rescaling comes from the shift orbifold itself. Such orbifolds always have divergent normalization constants. Here we are dealing with such constants within a specific regularization scheme.} These are such that
\be
\frac{\mu_{p}(\epsilon)}{2\sqrt{\epsilon}R(\epsilon)} = \mu_{p,\mathrm{phys}}\ .
\ee
Following the harmonic oscillator example, we will make the choice
\be
 R^2(\epsilon) = -\frac{1}{4\epsilon\ln(R^2_{\mathrm{phys}}\epsilon)} \ .
\ee

With these choices, the Lagrangian becomes
\be\label{eq:almost}
 \mathcal{L}_{W}= - \mu_{p}e^{-\varphi}\sqrt{g}\left(2+2\pi \alpha'|\partial\tau|^2+g_{11}(\partial\ell)^2 - |\tau|^2\left(\pi+\frac{2g_{11}R^2}{2\pi\alpha'}\cos\left(\frac{\ell}{R }\right)\right)\right) \ .
\ee
Equation (\ref{eq:almost}) is almost correct, but it does not yet have the correct $R\rightarrow\infty $ limit.  We must add a term which is the analog of the vacuum energy in (\ref{eq:vaccond}). Here the term induces a metric-dependent mass term to the $\tau$ field.

After fixing the counterterms we may state that the effect of the method of images is to replace $\frac{1}{2}\ell^2\rightarrow R^2\left(1-\cos(\frac{\ell}{R})\right)$,
\bea\label{eq:compactTDBI}
 \mathcal{L}_W &=& -2\mu_{p} e^{-\varphi}\sqrt{-g}- \ \mu_{p-1} e^{-\varphi} \sqrt{-g} \Big{(}g_{11}(\partial \ell)^2 \nonumber\\ &&\ \ \ \ \ \ \ \ \ \ \ \ \ \ \ +\ 2\pi \alpha' |\partial \tau|^2+g_{11}\frac{2|\tau|^2R^2\left(1-\cos(\frac{\ell}{R})\right)}{2\pi\alpha'}-\pi|\tau|^2\Big{)} \ .
\eea
The Lagrangian in (\ref{eq:compactTDBI}) still knows about the parameter $R$ as opposed to simply truncating to the zero winding sector. Because the effective potential is bounded from above as a function of $\ell$, generically we expect the backgrounds considered here to be energetically favorable when compared to a truncation to zero winding modes.  After T-duality, dependence on this length scale directly corresponds to inhomogeneities in the tachyon condensation. In the next Section \ref{sec:SS} we  apply the effective action (\ref{eq:compactTDBI}) to a model of chiral symmetry breaking in HQCD and explicitly show that it is energetically preferred.


\section{Application to HQCD model }\label{sec:SS}

In \cite{Sakai:2004cn,Sakai:2005yt} Sakai and Sugimoto introduced a D4-D8-$\overline{\textrm D}$8 brane system which provides a geometric realization of chiral symmetry breaking of holographic QCD.\footnote{In this Section we shall conform to the notations in references \cite{Sakai:2004cn,Sakai:2005yt}.  Specifically, the compact coordinate will be $x^4$ with radius $R_4$.  The parameter $R$ will refer to the AdS scale.  In addition, we will work in a gauge where the tachyon $\tau$ is real and define $\phi = |\tau|$.} In the following we shall only consider the geometric part of the SS model, for a discussion of the gauge field theory, see the review \cite{Erdmenger:2007cm}. In the limit of large number of D$4$-branes ($N_c\to\infty$) we can treat D$8$-$\overline{\textrm D}$8-branes as probes in this geometry. In particular, to demonstrate our methods we only consider a single ($N_f=1$) D$8$-$\overline{\textrm D}$8 pair.

The D$4$-brane near horizon geometry is
\be
 ds^2 = \left(\frac{\alpha'U}{R}\right)^{3/2}\left(\eta_{\mu\nu}dx^\mu dx^\nu
+f(U)(dx^4)^2\right)+\left(\frac{\alpha'U}{R}\right)^{-3/2} \left(\frac{(\alpha'dU)^2}{f(U)}+(\alpha'U)^2d\Omega^2\right)
\ee
with nontrivial dilaton and three-form
\be
e^{\varphi} = g_s\left(\frac{\alpha'U}{R}\right)^{3/4} \ , \ F_4 = dC_3 = \frac{2\pi N_c}{V_4}\epsilon_4 \ , \
 f(U) =  1-\left(\frac{U_{KK}^3}{U^3}\right) \ ,\ U_{KK} = \frac{4}{9}\frac{R^3}{R_4^2}\ .
\ee
The coordinate $U$ parameterizes the radial direction of $AdS_6$ of characteristic length scale $R$.  In this background the $x^4$-direction is compact with radius $R_4$. D8-$\overline{\textrm D}$8-branes are pointlike and separated by length $\ell$ in the $x^4$-direction.

We are interested in studying the (complex) tachyon in the SS model,
which originates from the open strings stretching between D8- and
$\overline{\textrm D}$8-branes. Basically, such a modification of the original model has
already been studied in \cite{Bergman:2007pm,Dhar:2007bz,Dhar:2008um}, but here we
want to keep the $x_4$-direction compact. The simplest thing, and the
background closest to the non-compact case, is to consider only the
zero winding tachyon. Naively, this seems like to be the only choice, but it
has some consequences. The first is that the chiral symmetry breaking
is completely independent of the size of the circle $R_4$, which in turn is related to the confinement
scale. The second is that the T-dual system could only have
homogeneous tachyon decay.  From other discussions of open string
tachyons \cite{Sen:2002vv,Larsen:2002wc,Lambert:2003zr,Balasubramanian:2004fz,Jokela:2005ha}, one would expect inhomogeneous decay to be relevant.  Here we consider a tachyon background with higher winding
modes turned on (the T-dual profile is highly inhomogeneous) and later
show that the higher winding profile is preferred when the branes are
macroscopically separated and the tachyon is ``small''.

For simplicity, we will make the following assumptions. We will consider a background of vanishing gauge field $C_3=0$. Also, the tachyon field and transverse scalar are only assumed to have spatial $U$-dependence, $\phi = \phi(U)$ and $\ell = \ell(U)$, as in \cite{Bergman:2007pm,Dhar:2007bz,Dhar:2008um}.

As before, we can expand the tachyon DBI action for the D8-$\overline{\textrm D}$8 system to second order.  Explicitly,
\be
g_{44} = f(U)\left(\frac{\alpha' U}{R}\right)^{3/2 }\ ,\ \sqrt{-g}  = \frac{\alpha'^5 U^4}{f(U)^{1/2}}\left(\frac{\alpha' U}{R}\right)^{-3/4}\ ,
\ee
and
\bea
 S_{NW} & = & -\mathcal N\int_{U_{KK}}^\infty  \frac{dU}{\sqrt{f(U)}}\left(\frac{\alpha'U}{R}\right)^{5/2}\Bigg\{ 1+\half f^2(U)\left(\frac{\alpha' U}{R}\right)^3(\pat_U \ell)^2  \label{eq:action}\\
        &   & +\pi\alpha'f(U)\left(\frac{\alpha' U}{R}\right)^{3/2}(\pat_U\phi)^2
+\left[-\half\pi + f(U)\left(\frac{\alpha' U}{R}\right)^{3/2}\frac{\ell^2}{4\pi\alpha'}\right]\phi^2 \Bigg\} \ ,\nonumber
\eea
where $\mathcal N = \frac{2V_{3+1}\Omega_4 \alpha'R^{4} \mu_8}{g_s}$ and subscript $NW$ indicates no winding.

The equations of motion derived from (\ref{eq:action}) are
\bea \label{eq:zweom}
 \pat_u\left( u^{11/2}f^{3/2}(u)(\pat_u \ell(u)) \right) & = & \frac{R^2}{2\pi \left(\alpha'\right)^3}u^4\sqrt{f(u)}\phi^2(u)\ell(u) \\
 \pat_u\left( u^4\sqrt{f(u)}(\pat_u \phi(u)) \right)   & = & \frac{R^2}{2\pi\left(\alpha'\right)^3} \left\{ -\frac{\pi u^{5/2}}{\sqrt{f(u)}}+\frac{u^4\sqrt{f(u)}\ell^2(u)}{2\pi\alpha'}  \right\}\phi(u) \nonumber \ ,
\eea
where $u=U/R$.
We label the solutions of these by $\phi_{NW}(u),\ell_{NW}(u)$.

\paragraph{Multiple winding case: }

The renormalized action (\ref{eq:compactTDBI}) for multiple winding tachyons in the D4-brane background reads
\bea
 S_{W} & = & -\mathcal N\int_{U_{KK}}^\infty  \frac{dU}{\sqrt{f(U)}}\left(\frac{\alpha'U}{R}\right)^{5/2}\Bigg\{ 1+\half f^2(U)\left(\frac{\alpha' U}{R}\right)^3(\pat_U \ell)^2 \label{eq:action2} \nonumber\\
       &   & +\ \  \pi\alpha'f(U)\left(\frac{\alpha' U}{R}\right)^{3/2}(\pat_U\phi)^2\nonumber\\
       &   & + \left[-\half \pi + f(U)\left(\frac{\alpha' U}{R}\right)^{3/2}\frac{2 R_4^2\left[1-\cos\left(\ell/R_4\right)\right]}{4\pi\alpha'}\right]\phi^2 \Bigg\}\ .
\eea

The equations of motion we get from (\ref{eq:action2}) are now
\bea\label{eq:mweom}
 \pat_u\left( u^{11/2}f^{3/2}(u)(\pat_u \ell(u)) \right) & = & \frac{R^2}{2\pi \left(\alpha'\right)^3}u^4\sqrt{f(u)}\phi^2(u)R_4\sin\left(\ell(u)/R_4\right)\nonumber \\
 \pat_u\left( u^4\sqrt{f(u)}(\pat_u \phi(u)) \right)     & = & \frac{R^2}{2\pi\left(\alpha'\right)^3} \Bigg\{ -\frac{\pi u^{5/2}}{\sqrt{f(u)}}\nonumber\\
       &   &+\frac{u^4\sqrt{f(u)}R_4^2 \left[1-\cos\left(\ell(u)/R_4\right)\right]}{\pi\alpha'}  \Bigg\}\phi(u) \ .
\eea
We label the solutions of these by $\phi_W(u),\ell_W(u)$.

\subsection{Analysis of the solutions}

As dictated by the expansion of the action, we analyze the solutions of (\ref{eq:zweom}) and (\ref{eq:mweom}) when $u$ is large.  In which case, $\phi(u)$, $\pat_u\phi(u)$, and $\pat_u \ell(u)$ are expected to be small, but $\ell(u)$ may be large. When these assumptions are violated a more detailed analysis is needed.
The infrared behavior of the nonexpanded (NW) action shows that $\ell(u)$ vanishes as $\phi(u)$ blows up (see \cite{Dhar:2008um,Bergman:2007pm}). Since the actions (\ref{eq:action}) and (\ref{eq:action2}) coincide for $\ell(u) \to 0$, we expect that winding effects do not change the infrared behavior essentially.

It is known \cite{Dhar:2008um,Bergman:2007pm} that
for $u \gg  1/\ell_\infty^{4/3}$ the ultraviolet normalizable solution is exponentially damped:
\be \label{eq:phiasympt}
 \phi_{NW}(u) \sim \frac{1}{(bu)^2} e^{-bu} \ ,
\ee
where $b=R \ell_\infty/(2\pi\alpha')$ and $\ell_\infty$ is the constant (asymptotic) value of $\ell(u)$. Notice that $\ell_\infty\leq \pi R_4$. The solution of $\ell_{NW}(u)$ 
sufficiently far away from the zero of $f(u)$ reads
\be
 \ell_{NW}(u) \simeq \ell_\infty -\frac{C}{u^{9/2}}
\ee
plus a term which is driven by the tachyon field. This additional term may be dominant for $u \sim 1/\ell_\infty^{4/3}$ if the tachyon field is large enough, but we shall not consider such cases. The asymptotic behavior of the solutions of (\ref{eq:mweom}), $\ell_W(u)$ and $\phi_W(u)$, are obtained from the above formulae by the replacement $\ell_\infty \to R_4\sqrt{2(1-\cos(\ell_\infty/R_4))}$ in (\ref{eq:phiasympt}).  This is simply seen by comparing \eq{eq:zweom} and \eq{eq:mweom}. 

We now turn to analyzing the full set of equations of motion (\ref{eq:zweom}) and (\ref{eq:mweom}) by numerical methods, in the ultraviolet regime $u\gg 1/\ell_\infty^{4/3}$.

\begin{figure}
\begin{center}
\noindent
\includegraphics[width=0.5\textwidth]{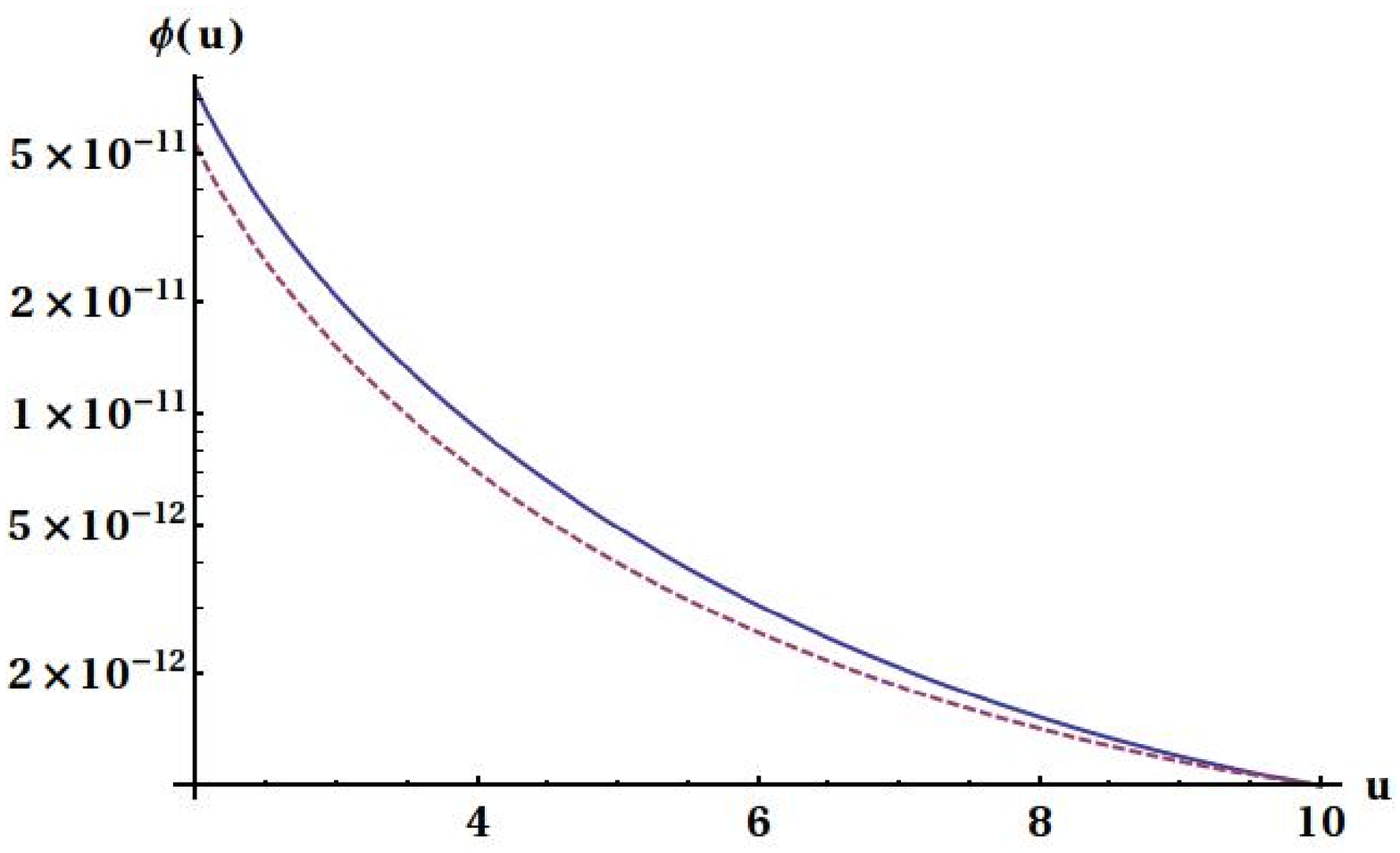}\includegraphics[width=0.5\textwidth]{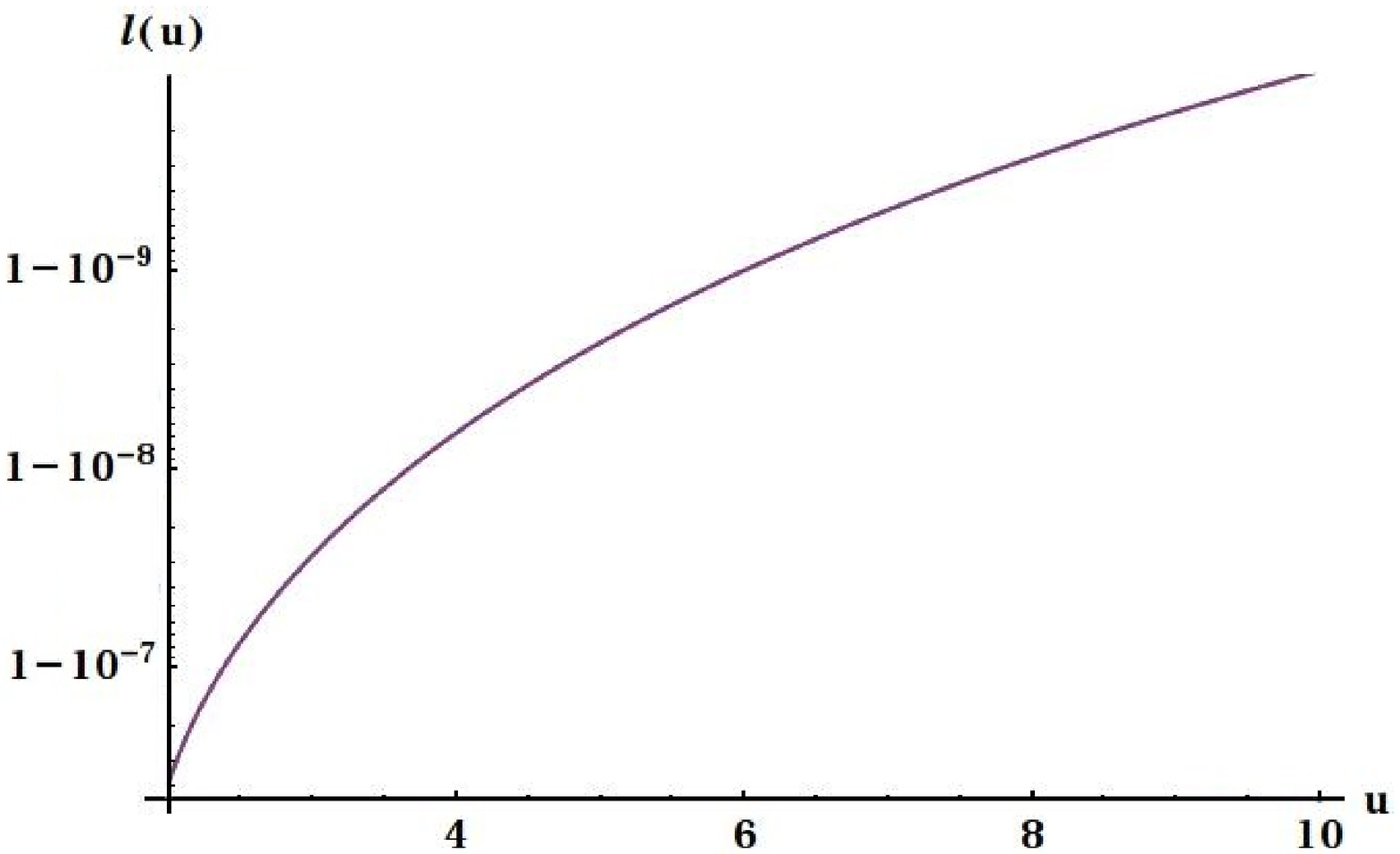}
\end{center}
\caption{The solutions of the differential equations (\ref{eq:zweom}) (solid curves) and (\ref{eq:mweom}) (dashed curves) for $\phi(u)$ (left) and $\ell(u)$ (right). The solutions for $\ell(u)$ overlap. Boundary conditions at $u=10$ are $\ell'(10) = 0.22\times 10^{-11}$, $\phi(10)=10^{-12}$ and $\phi'(10)=-10^{-12}/2\pi$. We use $R_4=1/2$ and $\ell_\infty=1$. \label{fig:philsol}}
\end{figure}

Let us first compare the classical solutions coming from the zero and the multiple winding actions in the ultraviolet region, \emph{i.e.}, in the region where $\ell(u)$ grows monotonically, its deviation from the asymptotic value is small, and the normalizable part of the tachyon decays exponentially. The $R$ dependence of the actions may be removed by rescaling $u \to u/R^4$, $\ell(u) \to R^3 \ell(u)$, and $R_4 \to R^3 R_4$ \cite{Dhar:2008um,Bergman:2007pm}. Remembering to rescale the boundary conditions, we will fix $R=\alpha' =1$. The solutions for specific boundary values at $u=10$, with $\ell'(10)\sim 10^{-11}$ and $\phi(10) \sim 10^{-12}$, are shown in Fig.~\ref{fig:philsol}.

\begin{figure}
\begin{center}
\noindent
\includegraphics[width=0.7\textwidth]{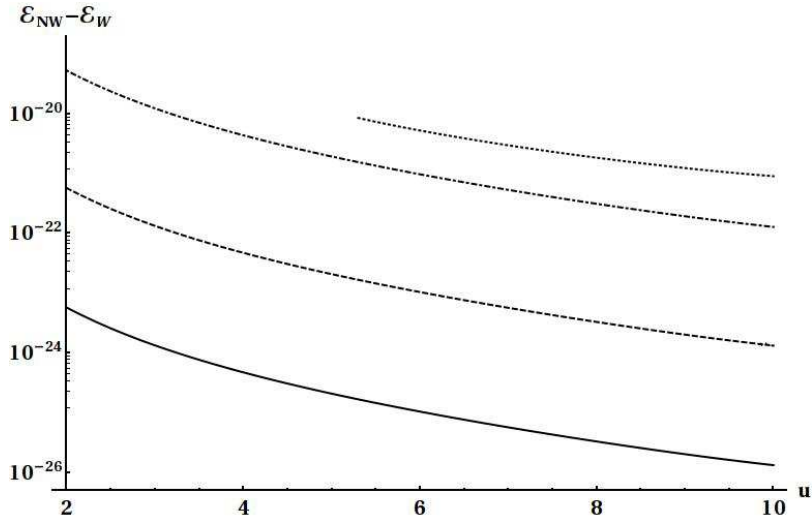}
\end{center}
\caption{The difference of the energy densities corresponding to the actions (\ref{eq:action}) and (\ref{eq:action2}). The boundary conditions of the fields are the same as in Fig.~\ref{fig:philsol}. We set $\ell_\infty=R={\cal N} =1$, and vary $R_4$. Solid, dashed, dot-dashed, and dotted curves correspond to $R_4 = 100,10,1$, and $1/\pi$, respectively. For the antipodal case, $R_4=1/\pi$, we restricted $u \gtrsim 5$ due to the zero of $f(u)$ at $u\simeq 4.4$.
\label{fig:Ediff}}
\end{figure}

Using the numeric solutions it is straightforward to evaluate the energy density difference $\mathcal E_{NW}-\mathcal E_W$ in the ultraviolet region. The difference of the energy densities of the two actions is plotted in Fig.~\ref{fig:Ediff} using the same boundary conditions for the fields as in Fig.~\ref{fig:philsol} but varying $R_4$.
It is seen that the qualitative behavior is stable against changes in $R_4$. We have also checked
that the action with multiple windings is slightly energetically favored\footnote{The IR (small $u$) physics does not change this result since in this case the on-shell actions of zero and multiple windings coincide.}
\be
 \mathcal{E}_W < \mathcal{E}_{NW} \ .
\ee
This is easy to understand by using the asymptotics \eq{eq:phiasympt}. When winding is included, the constant $b$ is multiplied by the factor
$R_4\sqrt{2(1-\cos(\ell_\infty/R_4))}/\ell_\infty$, which is always smaller than one. Thus the exponential decay is slower, leading to smaller values of the tachyon field as seen in Fig.~\ref{fig:philsol} (left). The tachyon contributions to the energy densities are positive ($\propto \phi'(u)^2$, $\phi(u)^2$) and hence smaller in the multiple winding case. The contributions to the difference $\mathcal E_{NW}-\mathcal E_W$  due to the change in $\ell(u)$ are subleading.

\section{Discussion}\label{sec:conc}

We developed a novel method to study classical theories of separated branes with compact transverse dimensions and inhomogeneous open string tachyon condensation along a circle. While describing such spacetimes with a shift orbifold, we introduced scaling solutions which enabled us to deduce an effective action with only a finite number of fields. This method was explicitly applied to a simple harmonic oscillator example and to a linearized description of brane/anti-brane pairs. In both examples the fundamental result was to replace $\frac{\ell^2}{2}$ with $R^2\left(1-\cos(\frac{\ell}{R})\right)$ such that the correct $R\rightarrow \infty$ limit is obtained.

In Section \ref{sec:SS}, we applied the scaling solution to a single D8-$\overline{\mathrm{D}}$8-pair which was used in \cite{Sakai:2004cn,Sakai:2005yt} to describe chiral symmetry breaking.  Where our analysis was valid, at large $U$, we found that the scaling solution has a lower free energy than simply truncating physics to the zero winding sector. At the very least, we take this to indicate that scaling solutions are competitive with the standard backgrounds used previously.

Given our apparent success in finding nontrivial scaling solutions to the low-energy effective action, we would like to demonstrate their existence at a more fundamental level. An incremental approach would be to improve the analysis in Subsection \ref{sec:TDBIcirc}, where we found it necessary to expand the square-root in the DBI action. When treating the higher powers order by order, we lose the DBI action's exactness in $\alpha'$. It would be instructive to try to deal with this action without performing such an expansion.  At a more fundamental level, one would like to move away from low-energy effective actions and directly deal with the worldsheet description. The challenge with trying to solve the worldsheet theory is that one must simultaneously turn on a nontrivial tachyon and length modulus backgrounds, a task which is currently out of reach.

In addition to identifying a scaling solution in the fundamental string theory, many questions relevant to HQCD arise from the results of Section \ref{sec:SS}. For example, to study chiral symmetry breaking we would like to generalize the discussion to include more flavors. Also, it would be interesting to incorporate finite temperature into the analysis. As in \cite{Edalati:2008xr}, it should also be possible to generalize to $p-q$ brane configurations and to non-vanishing worldvolume gauge fields.

Authors of \cite{Casero:2007ae,Bergman:2007pm} introduced the tachyon field, in the non-compact case, to study quark masses.  They
found Gell-Mann--Oakes--Renner (GOR) type relationship relating the quark expectation values to pion masses.
To do this it was necessary to use the full tachyon DBI action and impose boundary conditions on the tachyon when
the branes joined in the IR.  It would be very interesting to study the effect of a compact direction on the GOR
relationship.  Unfortunately, this is beyond the scope of our current analysis, which is only valid when the
branes are well separated.

\bigskip

\bigskip

\noindent

{\bf \large Acknowledgments}

\bigskip

We thank O. Bergman, M. Edalati, K. Hashimoto, P. Hoyer, S. Kawai, E. Keski-Vakkuri, P. Kraus, M. Lippert, R.G. Leigh, J. Majumder, C. Montonen, A. Sen, S. Terashima, and P. Yogendran for useful discussions. N.J. has been in part supported by the Magnus Ehrnrooth foundation and in part by the Israel Science Foundation under grant no. 568/05. M.J. has been in part supported by the Marie Curie Excellence Grant under contract MEXT-CT-2004-013510.  This work was also partially supported by the EU 6th Framework Marie Curie Research and Training network ``UniverseNet'' (MRTN-CT-2006-035863).

\bigskip

\bigskip


\appendix

\section{Analytic regularization}\label{app:anal}

It is interesting to study the Lagrangian in (\ref{sumosc}) using an alternate regulator.  The most obvious alternative, which preserves invariance under shifts in $\theta$, is an analytic continuation regulator
\bea
\mathcal{L}   & = & \lim_{s\rightarrow-1}\mathcal{L}_s \\
 \mathcal{L}_s & = & \left(\frac{m R^2}{2}\right)^{-s}\sum_{n=-\infty}^\infty\left(\frac{1}{\dot{\theta}^2+\frac{k}{m}(\theta+2\pi n)^2}\right)^{s} \ .
\eea

As before we want to introduce counterterms which preserve the shift symmetry, $\tilde{Z}_\ell = \tilde{Z}_R^2$. We find,
\be
 \mathcal{L}_{s} = \left(\frac{m \tilde{Z}_R^2R^2}{2}\right)^{-
s}\sum_{n=-\infty}^\infty\left(\frac{1}{\dot{\theta}^2+\frac{\tilde{Z}_k k}{m}(\theta+2\pi n)^2}\right)^{s} \ .
\ee
At this point we are implicitly thinking of the analytic counterterms, $\tilde{Z}_i$, as functions of the regulating parameter, $s$.

In actuality, it is simpler to introduce modified counterterms, $\bar{Z}_i(t)$, after introducing Schwinger parameters:
\bea
 \mathcal{L}_s & = & \left(\frac{m R^2}{2}\right)^{ - s}\frac{1}{\Gamma(s)}\int_0^\infty \frac{dt}{t}\ \bar{Z}_R^2t^{s}\sum_{n=-\infty}^\infty e^{-t\left( \dot{\theta}^2+\frac{\bar{Z}_k k}{m}(\theta+2\pi n)^2 \right)} \\
               & = & \left(\frac{m R^2}{2}\right)^{ - s}\frac{1}{\Gamma(s)}\int_0^\infty \frac{dt}{t}\ \bar{Z}_R^2t^s e^{-t(\dot{\theta}^2+\frac{\bar{Z}_k k}{m}\theta^2)}\Theta_3\left(2i\frac{\bar{Z}_kk}{m}t\theta\Big{|}4\pi i \frac{\bar{Z}_kk}{m}t\right) \\
               & = & \left(\frac{m R^2}{2}\right)^{- s}\!\!\frac{1}{\Gamma(s)}\sqrt{\frac{m}{4\pi k}}\int_0^\infty \frac{dt}{t}\ \frac{\bar{Z}_R^2}{\sqrt{\bar{Z}_k}} t^{s-1/2}e^{-t\dot{\theta}^2}\Theta_3\left(\frac{\theta}{2\pi}\Big{|}\frac{im}{4\pi \bar{Z}_k k t }\right) \\
               & = &  \left(\frac{m R^2}{2}\right)^{- s}\!\!\!\frac{1}{\Gamma(s)}\sqrt{\frac{m}{4\pi  k}}\!\int_0^\infty\! \frac{dt}{t}\frac{\bar{Z}_R^2}{\sqrt{\bar{Z}_k}} t^{s-1/2}e^{-t\dot{\theta}^2}\!\!\left(\!1\!+2\sum_{q=1}^\infty\cos(\theta q) e^{\frac{-1}{4t}\left(q\sqrt{\frac{m}{\bar{Z}_k k}}\right)^2}\right) \; . \nonumber
\eea
This integral expression serves as an implicit definition of the modified counterterms $\bar{Z}_i(t)$ in terms of  $\tilde{Z}_i(s)$.  Though this appears to be a complicated integral expression, we can solve for the $\bar{Z}_i(t)$.
\paragraph{{\bf Analytic counterterms:\\}}

It is a straightforward exercise to solve for the counterterms which are consistent with the renormalization conditions.  The solution is
\bea
 \bar{Z}_\ell(t) & = & \bar{Z}^2_R(t)\\
 \bar{Z}^4_R(t)  & = & \frac{4\pi k}{m}t\bar{Z}_k(t)\\
 \bar{Z}_k(t)    & = & -\frac{m}{4k}\frac{1}{t\ln\left(\frac{k}{m}t\right)}\\
 \bar{Z}_\Lambda & = & kR^2\\
 \mathcal{L}_s   & = & \frac{mR^2}{2\Gamma(s)}\sum_{q=-\infty}^\infty e^{iq\theta}\left(\frac{k}{m}\right)^{q^2}\dot{\theta}^{-(s+q^2)}\Gamma(s+q^2)+kR^2 \\
                 & \sim & \frac{m R^2}{2}\dot{\theta}^2+kR^2\left(1-\cos(\theta)\right) \ .
\eea

\paragraph{{\bf Mellin transformation:\\}}

Though the cut-off scheme and the analytic scheme look different, they are related through a Mellin transform,
\bea
 A(\epsilon)                             & \equiv & \sum_{n=-\infty}^\infty e^{-\lambda_n\epsilon} \ , \
 A(s)                                    \equiv \sum_{n=-\infty}^\infty \lambda_n^{-s} \\
 \mathcal{M}\left[A(\epsilon)\right](s)  &    =   & \int_0^\infty \frac{d\epsilon}{\epsilon} \epsilon^{s} A(\epsilon)
                                         = \Gamma(s) A(s)\ .
\eea
The most important fact to note is that, while a cut-off acts locally in momentum space, its effects are smeared out when defining the analytic scheme.  In essence, we are Fourier transforming the cut-off dependence.

While this is obviously true before introducing the scaling solution, knowledge of the counterterms for the analytic regulator allows us to check the Mellin transform relationship for the renormalized action.
\bea
\frac{\mathcal{M}[\mathcal{L}(\epsilon)](s)}{\Gamma(s)} &   =  & \frac{mR^2}{2\Gamma(s)}\int_0^\infty \frac{d\epsilon}{\epsilon} \epsilon^s\left(\sum_{q=-\infty}^\infty e^{iq\theta}\left(\frac{k}{m}\epsilon\right)^{q^2}\right)+Z_\Lambda(s) \\
                &   =  & \frac{mR^2}{2\Gamma(s)}\sum_{q=-\infty}^\infty e^{iq\theta}\left(\frac{k}{m}\right)^{q^2}\left(\dot{\theta}\right)^{-(s+q^2)}\Gamma(s+q^2)+ Z_\Lambda(s) \\
                & \sim & \frac{mR^2}{2}\dot{\theta}^2\frac{\Gamma(s)}{\Gamma(s)}+k R^2 \cos(\theta)\frac{\Gamma(s+1)}{\Gamma(s)} + Z_\Lambda(s) \ .
\eea
After setting $Z_\Lambda(s) = kR^2$, we obtain the correct effective Lagrangian\footnote{Note that $\lim_{\epsilon\rightarrow 0}\frac{\Gamma(-n+\epsilon)}{\Gamma(-(n+1)+\epsilon)} = -(n+1).$}
\be
 \mathcal{L}_s = \frac{mR^2}{2}\dot{\theta}^2 + kR^2\left(1-\cos(\theta)\right)+\mathcal{O}(s+1) \ .
\ee

In this Appendix we have dealt with the divergences in (\ref{sumosc}) using an alternate regulator.  We find that the renormalized effective action is the same for both regulators.


\begin{thebibliography}{99}

\bibitem{Sakai:2004cn}
  T.~Sakai and S.~Sugimoto,
  Prog.\ Theor.\ Phys.\  {\bf 113} (2005) 843
  [arXiv:hep-th/0412141].

\bibitem{Sakai:2005yt}
  T.~Sakai and S.~Sugimoto,
  Prog.\ Theor.\ Phys.\  {\bf 114} (2005) 1083
  [arXiv:hep-th/0507073].

\bibitem{Bergman:2007pm}
  O.~Bergman, S.~Seki and J.~Sonnenschein,
  JHEP {\bf 0712} (2007) 037
  [arXiv:0708.2839 [hep-th]].

\bibitem{Dhar:2007bz}
  A.~Dhar and P.~Nag,
  JHEP {\bf 0801} (2008) 055
  [arXiv:0708.3233 [hep-th]].

\bibitem{Dhar:2008um}
  A.~Dhar and P.~Nag,
  arXiv:0804.4807 [hep-th].

\bibitem{Taylor:1996ik}
  W.~Taylor,
  Phys.\ Lett.\  B {\bf 394} (1997) 283
  [arXiv:hep-th/9611042].

\bibitem{Sugimoto:2004mh}
  S.~Sugimoto and K.~Takahashi,
  JHEP {\bf 0404} (2004) 051
  [arXiv:hep-th/0403247].

\bibitem{Kraus:2000nj}
 P.~Kraus and F.~Larsen,
 Phys.\ Rev.\  D {\bf 63} (2001) 106004
 [arXiv:hep-th/0012198].

\bibitem{Takayanagi:2000rz}
 T.~Takayanagi, S.~Terashima and T.~Uesugi,
 JHEP {\bf 0103} (2001) 019
 [arXiv:hep-th/0012210].

\bibitem{Jones:2002sia}
 N.~T.~Jones and S.~H.~H.~Tye,
 JHEP {\bf 0301} (2003) 012
 [arXiv:hep-th/0211180].

\bibitem{Sen:2003tm}
 A.~Sen,
 Phys.\ Rev.\  D {\bf 68} (2003) 066008
 [arXiv:hep-th/0303057].

\bibitem{Sen:2004nf}
 A.~Sen,
 Int.\ J.\ Mod.\ Phys.\  A {\bf 20} (2005) 5513
 [arXiv:hep-th/0410103].

\bibitem{Garousi:2004rd}
 M.~R.~Garousi,
 JHEP {\bf 0501} (2005) 029
 [arXiv:hep-th/0411222].

\bibitem{Garousi:2007fn}
 M.~R.~Garousi,
 JHEP {\bf 0712} (2007) 089
 [arXiv:0710.5469 [hep-th]].

\bibitem{Terashima:2008jz}
  S.~Terashima,
  arXiv:0806.0975 [hep-th].

\bibitem{Sen:2003zf}
  A.~Sen,
  JHEP {\bf 0403} (2004) 070
  [arXiv:hep-th/0312003].

\bibitem{Sen:2002vv}
  A.~Sen,
  JHEP {\bf 0210} (2002) 003
  [arXiv:hep-th/0207105].

\bibitem{Larsen:2002wc}
  F.~Larsen, A.~Naqvi and S.~Terashima,
  JHEP {\bf 0302}, 039 (2003)
  [arXiv:hep-th/0212248].

\bibitem{Lambert:2003zr}
  N.~D.~Lambert, H.~Liu and J.~M.~Maldacena,
  JHEP {\bf 0703}, 014 (2007)
  [arXiv:hep-th/0303139].

\bibitem{Balasubramanian:2004fz}
  V.~Balasubramanian, E.~Keski-Vakkuri, P.~Kraus and A.~Naqvi,
  Commun.\ Math.\ Phys.\  {\bf 257}, 363 (2005)
  [arXiv:hep-th/0404039].

\bibitem{Jokela:2005ha}
  N.~Jokela, E.~Keski-Vakkuri and J.~Majumder,
  Phys.\ Rev.\  D {\bf 73}, 046007 (2006)
  [arXiv:hep-th/0510205].

\bibitem{Antonyan:2006vw}
  E.~Antonyan, J.~A.~Harvey, S.~Jensen and D.~Kutasov,
  arXiv:hep-th/0604017.

\bibitem{Aharony:2006da}
  O.~Aharony, J.~Sonnenschein and S.~Yankielowicz,
  Annals Phys.\  {\bf 322} (2007) 1420
  [arXiv:hep-th/0604161].

\bibitem{Parnachev:2006dn}
  A.~Parnachev and D.~A.~Sahakyan,
  Phys.\ Rev.\ Lett.\  {\bf 97} (2006) 111601
  [arXiv:hep-th/0604173].

\bibitem{Nawa:2006gv}
  K.~Nawa, H.~Suganuma and T.~Kojo,
  Phys.\ Rev.\  D {\bf 75} (2007) 086003
  [arXiv:hep-th/0612187].

\bibitem{Bergman:2006xn}
  O.~Bergman and G.~Lifschytz,
  JHEP {\bf 0704} (2007) 043
  [arXiv:hep-th/0612289].

\bibitem{Hata:2007mb}
  H.~Hata, T.~Sakai, S.~Sugimoto and S.~Yamato,
  Prog.\ Theor.\ Phys.\  {\bf 117} (2007) 1157
  [arXiv:hep-th/0701280].

\bibitem{Casero:2007ae}
  R.~Casero, E.~Kiritsis and A.~Paredes,
  Nucl.\ Phys.\  B {\bf 787} (2007) 98
  [arXiv:hep-th/0702155].

\bibitem{Bergman:2007wp}
  O.~Bergman, G.~Lifschytz and M.~Lippert,
  JHEP {\bf 0711} (2007) 056
  [arXiv:0708.0326 [hep-th]].

\bibitem{Davis:2007ka}
  J.~L.~Davis, M.~Gutperle, P.~Kraus and I.~Sachs,
  JHEP {\bf 0710} (2007) 049
  [arXiv:0708.0589 [hep-th]].


\bibitem{Erdmenger:2007cm}
  J.~Erdmenger, N.~Evans, I.~Kirsch and E.~Threlfall,
  Eur.\ Phys.\ J.\  A {\bf 35} (2008) 81
  [arXiv:0711.4467 [hep-th]].

\bibitem{Bergman:2008sg}
  O.~Bergman, G.~Lifschytz and M.~Lippert,
  JHEP {\bf 0805} (2008) 007
  [arXiv:0802.3720 [hep-th]].

\bibitem{Johnson:2008vna}
  C.~V.~Johnson and A.~Kundu,
  JHEP {\bf 0812} (2008) 053
  [arXiv:0803.0038 [hep-th]].


\bibitem{Edalati:2008xr}
  M.~Edalati, R.~G.~Leigh and N.~N.~Hoang,
  JHEP {\bf 0905} (2009) 035
  [arXiv:0803.1277 [hep-th]].

\bibitem{Aharony:2008an}
  O.~Aharony and D.~Kutasov,
  Phys.\ Rev.\  D {\bf 78} (2008) 026005
  [arXiv:0803.3547 [hep-th]].

\bibitem{Hashimoto:2008sr}
  K.~Hashimoto, T.~Hirayama, F.~L.~Lin and H.~U.~Yee,
  JHEP {\bf 0807} (2008) 089
  [arXiv:0803.4192 [hep-th]].


\bibitem{Bergman:2008qv}
  O.~Bergman, G.~Lifschytz and M.~Lippert,
  arXiv:0806.0366 [hep-th].

\bibitem{Thompson:2008qw}
  E.~G.~Thompson and D.~T.~Son,
  arXiv:0806.0367 [hep-th].

\bibitem{Hashimoto:2008zw}
  K.~Hashimoto, T.~Sakai and S.~Sugimoto,
  arXiv:0806.3122 [hep-th].

\bibitem{McNees:2008km}
  R.~McNees, R.~C.~Myers and A.~Sinha,
  arXiv:0807.5127 [hep-th].

\bibitem{Seki:2008mu}
  S.~Seki and J.~Sonnenschein,
  arXiv:0810.1633 [hep-th].

\bibitem{Argyres:2008sw}
  P.~C.~Argyres, M.~Edalati, R.~G.~Leigh and J.~F.~Vazquez-Poritz,
  arXiv:0811.4617 [hep-th].




\end{thebibliography}
\end{document}